\documentclass[12pt]{article}
\usepackage{amsthm}

\newtheorem{Le}{Lemma}

\begin{document}

\title{Proof of Tightness of Varshamov - Gilbert Bound for Linear Binary Codes} 
\date{}
\author{Vladimir Blinovsky}
\date{\small
 Institute for Information Transmission Problems, \\
 B. Karetnyi 19, Moscow, Russia,\\
vblinovs@yandex.ru}
\maketitle\bigskip
\begin{center}

{\bf Abstract}
We find the asymptotically tight coding bound for linear codes.
We prove that asymptotic Varshamov - Gilbert~\cite{0},\ \cite{00} bound for linear binary codes is tight
\end{center}

In the linear space of $q-$ary vectors $(q=p^m ,\ p prime)$ of length $n$ as $F^n_q$, where alphabet is the set of elements from finite field $F_{q=p^m}$. Linear code $M_n$ is linear subspace of $F_q^n$,\ $k=\log_q M_n,\ R_{M_n}=k/n,\ d_{M_n}=\{ \min d(x,y),\ x\neq y\subset\} ,\ d(x,y)=\sum_{i=1}^n \delta{x_i ,y_i},\  \delta_{M_n} =d_{M_n}/n$, 

(Asymptotic) Varshamov -Gilbert bound~\cite{1},~\cite{2} says that there exist sequence of linear codes $F_q^n$ such that
\begin{equation}
\label{e12}
R_{M_n} >1-H_q(\delta_{M_n})+o(1) .
\end{equation}
as $n\to\infty$, where $H_q (x)=-x\log_q x -(1-x)\log_q (1-x)+(q-1)\log_q (x)$ is $q-$ary entropy. 

Main hypothesis in Information theory says that this bound is tight for binary alphabet. In present paper we prove this hypothesis and moreover find the tight asymptotic bound in general case of $q-$ary alphabet. 

We need the following Lemma~\cite{2}.
\begin{Le}
(StefanescuÕs, 2005). Let $p(x)\in R[x]$ be such that the number of variations of signs of its coefficients is even. If
$$
p(x)=c_1 x^{d_1}-b_1 x^{m_1}+c_2 x^{d_2}-b_2 x^{m_2} +\ldots + c_k x^{d_k}-b_k x^{m_k}+g(x),
$$
$g(x)\in R_{+}[x],\ c_i >0,\ b_i >0,\ D_i >m_i >d_{i+1}$ for all $i$, the number
$$
B_3 (p)=\max\left\{ \left(\frac{{b}_1}{c_1}\right)^{1/(d_1 -m_1)},\ldots ,\left(\frac{{b}_k}{c_k}\right)^{1/(d_k -m_k )}\right\} 
$$
is upper bound bound for the positive roots of the polynomial $p$ for any choice $c_1 ,\ldots ,c_k$.
\end{Le}
Define redundancy $r=n-k_{M_n}$ and $H_{M_n} = \{ h_{i,j},\ i\in [n],\ j\in [r]\}$ is checking matrix of the code $M_n$.

Let $h(i)$ be the rows of $H_{M_n}$.  
Consider polynome
\begin{equation}
\label{e1}
P_q (r,d)=\sum_{h (i)\in F^n_q} \prod_{\sigma\in \cup_{i=1}^{d-1} {[n]\choose k}_q}\left(q^r- q^{\sum_{j=1}^r \sum_{s=1}^m \cos  \left(\frac {2\pi}{p}2\pi  (h(j),\sigma)_s\right)}\right) .
\end{equation}
It is clear that this polynome is equal to zero iff $d_{M_n} <d$. Hence if we find the proper lower bound for the root $1/q^r$ of this polynome,
it would be the bound for the rate of the code $M_n$.

Expansion of $P_q (r,d)$ is as follows
\begin{eqnarray*}&&
P_q (r,d)=q^{r(V-1)}\sum_{j=0}^{V-1}(-1)^j  \frac{1}{q^{rj}}\sum_{\pi (j),\ \sum_s {k_s \leq r}}\frac{1}{\prod_{s=1}^j k_s !^2 s!^{k_{s}}} \\
&&\sum_{\{a_{m,\lambda}\} ,\{\sigma_{w}\}=\cup_{i=1}^{d-1} {[n]\choose k}_q }q^{\sum_{m=1}^j\sum_{\lambda =1}^{k_m}\sum_{t=1}^s \cos\left(\frac{2\pi}{p}(a_{m,\lambda}\sigma_{w_t})\right)} 
\end{eqnarray*}
where
$\{ a_{m,\lambda}\}$ runs over the set of rows of parity checking matrix. Using Lemma we find upper bound for the root $1/q^r$ of this polynome:
\begin{eqnarray}
\label{5t}
\frac{1}{q^r}\leq \max\left\{ (2s+2)(k_{2s+2})^2/((2s+1)(k_{2s+1})^2))\sum_\sigma q^{\sum_{s=1}^m \cos \left(\frac{2\pi}{p}(a,\sigma)_s\right)}\right\}.
\end{eqnarray}
Here $(a,\sigma)_s$ define $s-$th coordinate in the representation of the element $(a,\sigma)\in F_{p^m}$. 

Because $\max$ is taken over the set of numbers which are in logarithmic asymptotic equal, we have that bound~(\ref{5t}) is tight.
In binary case it gives (asymptotically) Varshamov- Gilbert bound.

Method using here -constructing of function which (positive) minimal root gives the upper bound for parameter of combinatorial 
problem and consequently finding the bound for this root  is original and we call it BlinovskyÕs (:)) averaging method.  We expect that this method allows to solve many combinatorial problems such as Hadwiger conjecture~\cite{3}, Borsuk conjecture~\cite{4}, codes with fixed distance $d$ and codes in Euclidean space. We hope we return to this scheme of proof in the subsequent manuscripts.

\end{document}